\documentstyle[12pt]{article}
\begin{document}

\begin{flushright}{\bf\underline{ Nov/1997}}\end{flushright}

\vspace{2cm}
\begin{center}
{\huge \bf The Compatibility between the Higher Dimensions Self Duality
and the Yang-Mills Equation of Motion}

\vspace{3.5cm}

{\large \bf Khaled Abdel-Khalek
\footnote{Work supported by an
ICSC--World Laboratory scholarship.\\
 e-mail : khaled@le.infn.it}
}
%\addtocounter{footnote}{-2}\\
\vspace{.4cm}

\end{center}

\vspace{3.5cm}

\begin{abstract}
We study the compatiblity between the higher dimension dualities and the
Yang-Mills equation of motion. Taking a 't~Hooft solution as a starting
point, we come to the conclusion that for only 4 dimensions the self
duality implies the equation of motion for generic instanton size.
Whereas in higher dimensions, the self duality is compatable with
the equation of motion, approximately, for small instanton size
i.e. the zero curvature condition. At the mathematical level, the self
duality is still useful since it transforms a second order
 into a first order  differential equation.
\end{abstract}

\newpage

\vspace{0.5cm}

Duality play an important role in physics. It represent weak form
of integrablity. In 4 dimensions, the Yang-Mills self dual solution,
instanton\cite{yy1}, is a non-perturbative notion.
Extending it to higher dimensions \cite{yy2,yy3,yy4,yy5,yy6}
is very important for studying p-branes
solitons\cite{yy7,yy8,yy9,yy10,yy11,yy12}
In recent article,
we proposed a new formulation of the 8 dimensional octonionic instanton
\cite{yy13}
which mimicks many of the quaternionic 4 dimensional instanton.
Our starting point was the construction of a Clifford algebra, Cliff(0,7),
from octonions
\begin{equation}
(E_i)_{\alpha\beta} = \delta_{i\alpha} \delta_{\beta 0} - \delta_{i \beta}
\delta_{\alpha 0} + f_{i\alpha\beta} ,
\end{equation}
$f_{ijk}$ is the octonionic structure constant. Noticing its
similarity with the 't~Hooft matrices.
Writing their commutation relations as
\begin{eqnarray}\label{as3}
[E_i , E_j]  &=&  2f_{ijk}
E_k -  2[ E_i, 1|E_j ] ,\\
&=& 2(f_{ijk} 1_{8\times 8} + [E_i, 1|E_j] E_k ) E_k , \label{lop}\\
&=& 2\rho_{ijk} E_k  .
\end{eqnarray}
So our structure constants become matrices\footnote{There is no
summation in the second term of (\ref{lop}) to avoid an ugly
$1/7$ factor.} and our varying torsion over $S^7$
is
\begin{equation}\label{tor}
T(X,Y) = - 2\rho_{ijk}.
\end{equation}
Now, introducing
\begin{equation}
E_\mu \equiv ( E_0 = 1 , E_i) \quad ; \quad 
\bar{E}_\mu \equiv (E_0, -E_i) ,
\end{equation}
we may define
 the following
tensor
\begin{equation}
\vartheta_{\mu\nu} =  {\frac{1}{2}}
( \bar{E}_\mu E_\nu - \bar{E}_\nu E_\mu ).
\end{equation}
which is self dual with respect to
\begin{equation}
\eta_{0ijk} = \rho_{ijk} \quad \mbox{and zero elsewhere},
\end{equation}
explicitly, we have
\begin{equation}
\vartheta_{\alpha \beta } =  \eta_{\alpha\beta\mu\nu}\vartheta_{\mu\nu}
\end{equation}
Then our 't~Hooft solution is 
\begin{equation}
A_\mu(x) = \frac{x^2 }{x^2  + \lambda^2}
g^{-1}(x) \partial_\mu g(x) = -
\frac{\vartheta_{\mu \nu} x^\nu}{\lambda^2 + x^2}
\end{equation}
for
\begin{equation}
F_{\mu\nu} =
\frac{\vartheta_{\mu \nu} 2 \lambda^2}{(\lambda^2 + x^2)^2} ,
\end{equation}
and our self-duality is
\begin{equation}\label{a3} 
F_{\alpha\beta} = 
\eta_{\alpha\beta\mu\nu} F_{\mu \nu} 
\end{equation} 
But the problem is in 4 dimensions, the self duality has
a very deep geometrical meaning. It is equivalent to the Bianchi
identities. So a self dual solution is automatically a solution
of the equation
of motion. Here, our solution does not satisfy the Yang-Mills
equation of motion
\begin{equation}
D_\mu F_{\mu\nu} = \partial_\mu F_{\mu \nu} +  [ A_\mu, F_{ \mu \nu } ] \neq 0,
\end{equation}
actually it comes too close
\begin{equation}\label{ll1}
3 \partial_\mu F_{\mu \nu} +  [ A_\mu, F_{ \mu \nu } ] \neq 0 .
\end{equation}

In this letter, we would like to know the reason of
this semilarity and if it possible to investigate the possibility
of having higher dimensions duality that is equivalent to the
equation of motion.

Rewriting
a generic self dual $A_\mu$ as
\begin{equation}
A_\mu(x) = \frac{x^2 }{(x^2  + \lambda^2)^m}
g^{-1}(x) \partial_\mu g(x) = a
\frac{G_{\mu \nu} x^\nu}{(\lambda^2 + x^2)^m}
\end{equation}
for a generic self dual matrix - to be found - and imposing 
\begin{equation}
F_{\mu\nu} = b
\frac{G_{\mu \nu} }{(\lambda^2 + x^2)^n} ,
\end{equation}
in order to get a  self-duality as
\begin{equation} 
F_{\alpha\beta} = 
\eta_{\alpha\beta\mu\nu} F_{\mu \nu}, 
\end{equation} 
we find that
\begin{equation}
m = 1 \quad n = 2 ,
\end{equation}
and
\begin{eqnarray}
b G_{\mu \nu} = -2 a (\lambda^2 + x^2) G_{\mu \nu }
+ 2 a (G_{\mu \alpha } x_\alpha x_\nu  -
G_{\nu \alpha } x_\alpha  x_\mu) +
a^2 x_\alpha  x_\beta  [ G_{\mu \alpha }, G_{\nu \beta } ] ,
\end{eqnarray}
leading to
\begin{equation}
b = - 2 \lambda^2
\end{equation}
and
\begin{equation}\label{f2}
(G_{\mu \alpha } \delta _{\nu\beta }  - G_{\nu \alpha }
 \delta _{\mu\beta }  ) +
\frac{a}{2}  [ G_{\mu \alpha }, G_{\nu \beta } ]  =
\delta_{ \alpha \beta}  G_{\mu \nu}.
\end{equation}
Whereas, taking $A_\mu$ and $F_{\mu\nu}$ and substituting
in the YM equation of motion,
\begin{equation}
\partial_\mu F_{\mu \nu} +  [ A_\mu, F_{ \mu \nu } ] =  0,
\end{equation}
we find
\begin{equation}\label{f1}
-4 G_{\alpha \nu } + a [ G_{\mu \alpha }, G_{\mu \nu } ] = 0.
\end{equation}
Now, we should find a common solution ($G_{\mu \nu }$) that satisfy
both of (\ref{f2}) and (\ref{f1}). We can reformulate
(\ref{f2}) for $\alpha = \beta $ then
\begin{equation}
(4 -2 dim) G_{\alpha \nu } + a [ G_{\mu \alpha }, G_{\mu \nu } ] = 0,
\end{equation}
 i.e (\ref{f2}) and (\ref{f1}) have a common 't~Hooft-like solution if
 and only if
 \begin{equation}\label{ll2}
(4 -2 dim) = -4 \quad \Longrightarrow \quad dim = 4.
 \end{equation}
 It is very easy to derive (\ref{ll1}) directly from (\ref{ll2}).
From the mathematical point of view, we have acheived some
success since we transformed a second order differential equation's
problem (\ref{ll1}) into the simple self duality relation (\ref{a3})
but physically, the self duality gives the YM equation of
motion
only for $\lambda \longrightarrow 0$ i.e. for very small instanton
size. One can  use some non-linear perturbation methods to construct
other ``acceptable'' solutions.

Let's try to find another self dual tensor and a new self duality in
8 dimensions and see what happens. We can define a new tensor
that plays the role of $\rho_{ijk}$ directly from
octonions. We can use $f_{ijk}$ the octonionic structure
constant and define
\begin{equation}
\pi_{0ijk} = f_{ijk} \quad \mbox{and zero elsewhere}
\end{equation}
our self dual tensor may be
\begin{equation}
\sigma_{\mu \nu } = \vartheta_{\mu \nu} - [ E_\mu, E_\nu ] ,
\end{equation}
leading to
\begin{equation}
\sigma_{\mu \nu } = \pi_{\mu \nu \alpha \beta } \sigma_{ \alpha \beta } .
\end{equation}
But, as one can see easily, this $\sigma_{\mu \nu} $ does not satisfy
(\ref{f2}).  
So, the self duality does not hold for $F_{\mu\nu}$.
It is not easy to guess a matrix that can
satisfy a self duality and the YM equation of motion at
the same time.

In summary, there is two ways, either to modify the 't~Hooft like
solution or to change the self duality formulation.
We prefer to choose the second way. We have a Cliff(0,7) structure
in our formulation so
\begin{equation}
E_i E_j = \rho_{ijk} E_k \Longleftrightarrow
E_i E_j = constant\ \ \epsilon_{ijlmnpk} E_k ,
\end{equation}
and in 8 dimensions, we have the standard
$\epsilon_{\alpha \beta \gamma \delta \zeta \xi  \mu \nu}$
Levi-Cevita tensor.
Having such tools, in principle, may be useful to study
a new type of gravitational instanton
\begin{equation}
R_{\alpha \beta \gamma \delta} =
\epsilon_{\alpha \beta \gamma \delta \zeta \xi  \mu \nu}
R_{\zeta \xi  \mu \nu}.
\end{equation}
where $R$ is our standard Reimanian tensor but may be formed
from a torsionful connection. Such point is under investigation.

\vspace{3.5 cm}

\section*{Acknowledgements}

I would like to thank P.~Rotelli for useful discusions about
ring division algebras.
Also, It is a pleasure to acknowledge  Prof. A.~Zichichi 
and the ICSC--World Laboratory for financial
support.

\vspace{3cm}
\begin{flushright}
{\bf
~~~~~Khaled Abdel-Khalek~~~~~~~~~~~~~ \\
~~~~~Dipartimento di Fisica \&~~~~~~~~~~ \\
Istituto Nazionale di Fisica Nucleare\\
~~~~~- Universit\`a di Lecce -~~~~~~~~~~~\\
~~~~~- Lecce, 73100, Italy -~~~~~~~~~~~}
\end{flushright}
\end{document}